\documentclass[%
aps,
prl,
reprint,
showpacs,
amsmath,amssymb,
]{revtex4-1}

\usepackage{graphicx}
\usepackage{dcolumn}
\usepackage{bm}
\usepackage[T1]{fontenc}
\usepackage{color}

\usepackage{amsmath}
\usepackage{amssymb}
\usepackage{esint}
\DeclareMathAlphabet{\mathpzc}{OT1}{pzc}{m}{it}

\newcommand{\ba}{\begin{align}}
\newcommand{\ea}{\end{align}}
\newcommand{\nn}{\nonumber}
\newcommand{\be}{\begin{equation}}
\newcommand{\ee}{\end{equation}} 
\newcommand{\bea}{\begin{eqnarray}}
\newcommand{\eea}{\end{eqnarray}}
\def\nn{\nonumber}

\begin{document}

\title{Rheological properties  of a dilute suspension of self-propelled particles}
\author{Moslem Moradi, Ali Najafi}
\affiliation{Physics Department, University of Zanjan, Zanjan 45371-38791, Iran}
\date{\today}

\begin{abstract}
With a detail microscopic model for a self-propelled swimmer, we derive the rheological 
properties of a dilute suspension of such particles at small Peclet numbers.
It is shown that, in addition to the Einstein's like contribution to the effective viscosity, that is proportional to  the volume fraction of the swimmers, a contribution due to the activity of self-propelled particles influences the viscosity. As a result of the activity of swimmers, the effective viscosity would be lower (higher) than the viscosity of the suspending medium when the particles are pushers (pullers).
Such activity dependent contribution, will also results to a non-Newtonian behavior of the  suspension in the form of normal stress differences. 
\end{abstract}
\pacs{87.16.-b,05.65.+b}
\maketitle
   
Dynamical properties of bacterial suspensions and suspensions of artificially designed 
self-propelled  micro particles have been the subject of many recent experimental 
and theoretical investigations \cite{Brennen,Paxton,Dreyfus,Ramin}. 
In addition to the self organized behavior which have been observed in such active suspensions 
\cite{Wu,Saintillan1,Dombrowski,Baskaran1,Vicsek,Baskaran2,Kruse,Liverpool}, 
it is an essential task to understand how such active suspensions response to  external 
forces and what rheological behavior they have \cite{Marchetti,Hatwalne,Foffano}.
Rheology  of active matter composed of self propelled particles is important and interesting from a fundamental point of view as in such systems the particles inject mechanical energy 
to the ambient fluid without applying any net hydrodynamical forces. 
Understanding the physics behind such phenomena is important for  microfluidic experiments that 
manipulate samples of microorganisms and also it could be relevant for micro-robots that are 
artificially developed.

Among all macroscopic rheological parameters of the system, the effective viscosity of such complex fluids is the main core of current investigations \cite{Saintillan2,Ishikawa}. It is a known experimental fact that the effective viscosity shows different behavior for active suspensions containing swimmers which their motion generated by head (puller) or tail (pusher). Recent experiment on suspension of motile algae $\mathit{Clomydomonas}$, shows that puller particles increase the effective viscosity~\cite{Rafai}. The effect of pusher particles are examined in an experiment performed on bacterium $\mathit{Bacillus\, subtilis}$~\cite{Sokolov1,Sokolov2}. It is shown that at small volume fraction of swimmers, the effective viscosity is smaller than the viscosity of ambient fluid, but at large volume fraction, the viscosity would be larger than the bare viscosity. In another experiment on $\mathit{E. coli}$, it is shown that at small Peclet number, the effective viscosity of pushers is smaller than the bare viscosity of the fluid~\cite{Gachelin}.    

Most of the theoritical works have been down so far, are theories with phenomenological origin~\cite{Hatwalne, Saintillan2}. In this letter, we use a microscopic model for pusher and puller particles and investigate their influence on the rheology of the suspension. This kind of description will allow us to have a detail insight on the role of microscopic parameters of the swimmer in the rheological properties.

\begin{figure} 
\includegraphics[width=1\columnwidth]{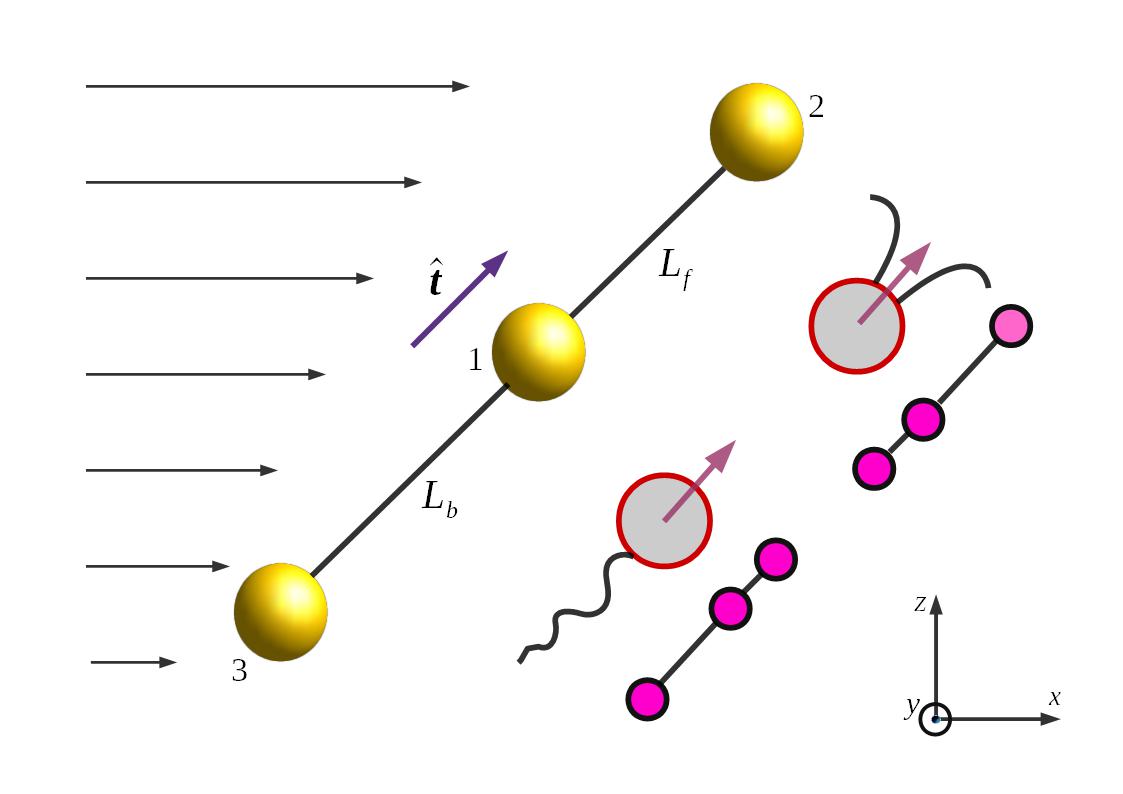}
\caption{Geometry of a low Reynolds swimmer immersed in a shear flow. 
For a non-symmetric swimmer where the back and front arms have different sizes, this system can model 
both pushers and pullers. }
\label{fig1}
\end{figure}

To study the rheological properties of a dilute suspension of microswimmers, we first study the dynamics of a single swimmer suspended in a Newtonian fluid with viscosity $\eta_{0}$ that is subject to an external shear. This will allow us to eventually achieve the response of the fluid to the external forces. As a mathematical model, there-sphere swimmer can model the general 
hydrodynamical  characteristics of most low Reynolds swimmers \cite{Najafi,Miri} and it captures the 
peculiarities at low Reynolds  hydrodynamics \cite{Taylor,Purcell,Najafi2}. Fig.~(\ref{fig1}) shows a schematic view of a three-sphere swimmer immersed in a purely straining shear flow. The swimmer consists of three spheres with radii  $a$ linked by two  front and back arms with variable  lengths given by ${L}_{f}(t) = L + u_{1}(t)$ and ${L}_{b}(t)= (1+\delta)L + u_{2}(t)$.  Here 
$L$ and $(1+\delta)L$, denote the mean arm lengths and $u_{1}(t)$ and $u_{2}(t)$ are two periodic functions of time with the same amplitude $u_0$. We further assume that these periodic functions average to zero. By considering an asymmetry parameter $\delta$, we will show how the asymmetry of the swimmer is essential in the overall dynamics of the system. The instantaneous orientation of the swimmer is denoted by a unit vector $\mathbf{\hat{t}}$ that  points from the swimmer's back to its front side. In terms of  polar variables $\theta$ and $\varphi$, this vector can be written as: 
$\mathbf{\hat{t}}=(\sin\theta\cos\varphi,\sin\theta\sin\varphi,\cos\theta)$. The externally applied shear flow is given by: ${\bf u}^{s}({\bf r})=\Gamma\cdot {\bf r}$, with $\Gamma=\dot{\gamma} \, \hat{\mathbf{z}} \, \hat{\mathbf{x}}$ where $\dot{\gamma}$ represents the shear rate. 
In the presence of the swimmer, the velocity filed of the fluid will change.  
To express the velocity field in the presence of the swimmer, and for mathematical tractability of the 
system  we assume that  sphere's diameter $a$ is much less than the arm's lengths ($a\ll L$). 
This assumption  allows us 
to consider the spheres as point forces.  Denoting the force exerted on fluid from the $j$'th sphere 
that is located at position ${\bf r}_j$ by ${\mathbf{f}}_{j}$, the velocity filed at a given point ${\bf r}$ reads as: 
\be
 \mathbf{u}(\mathbf{r}) =\sum_{j=1}^{3} {\cal O}(\mathbf{r} - {\mathbf{r}}_{j})\cdot {\mathbf{f}}_{j} + \Gamma \cdot \mathbf{r} + \mathbf{C}(\mathbf{r}) : \Gamma ,       
 \ee 
 where ${\cal O}(\mathbf{r}) =  (\frac{1}{8\pi{\eta}_{0} r})(\mathbf{I}+\hat{\mathbf{r}}\hat{\mathbf{r}})$ is the Green's function of the Stokes equation~\cite{Oseen}. Note that the Green's function regularizes to ${\cal O}(0) = (\frac{1}{6\pi\eta a}) \mathbf{I}$. The third rank disturbance tensor $\mathbf{C}$ represents the hydrodynamic interaction of two spheres in a shear flow and for sphere $i$ this disturbance matrix is given by~\cite{Dhont}:
\begin{align}\nn
\mathbf{C} ({\mathbf{r}}_{i})& =\sum_{j\neq i} \left( -\frac{5}{2} \frac{a^3}{{|{\mathbf{r}}_{ij}|}^3}+\frac{20}{3} \frac{a^5}{{|{\mathbf{r}}_{ij}|}^5} \right) \frac{{\mathbf{r}}_{ij}{\mathbf{r}}_{ij}{\mathbf{r}}_{ij}}{{|{\mathbf{r}}_{ij}|}^2}
\nonumber \\
&\quad - \frac{4}{3} \frac{a^5}{{|{\mathbf{r}}_{ij}|}^5} \left[   \mathbf{I} {\mathbf{r}}_{ij}+{(\mathbf{I} {\mathbf{r}}_{ij})}^{\dagger}\right] - \frac{25}{2} \frac{a^6}{{|{\mathbf{r}}_{ij}|}^6} \frac{{\mathbf{r}}_{ij}{\mathbf{r}}_{ij}{\mathbf{r}}_{ij}}{{|{\mathbf{r}}_{ij}|}^2}.
\end{align}
where ${\mathbf{r}}_{ij}={\mathbf{r}}_{i}-{\mathbf{r}}_{j}$, and ${\hat{\mathbf{r}}}_{ij}= ({\mathbf{r}}_{i} - {\mathbf{r}}_{j})/|{\mathbf{r}}_{i} - {\mathbf{r}}_{j}|$.
We are seeking the autonomous net swimming motion of the system. Denoting the velocity of 
$i$'th sphere by ${\bf v}_i$, the no slip boundary condition on the spheres requires that $\mathbf{u}({\mathbf{r}}_{i}) = {\mathbf{v}}_{i}$.  Internal dynamics of the arms are imposed by  
the following geometrical constraints:
\be 
{\mathbf{r}}_{2}-{\mathbf{r}}_{1}=L_{f}(t) \,\mathbf{\hat{t}}, \qquad {\mathbf{r}}_{1}-{\mathbf{r}}_{3}=L_{b}(t) \, \mathbf{\hat{t}}.
\ee
As there is no externally applied force or torque on the swimmer, the system should be force and torque 
free in the sense that  ${\mathbf{f}}_{1}+{\mathbf{f}}_{2}+{\mathbf{f}}_{3}=0$ 
and ${\bf r}_{21}\times{\bf f}_2+{\bf r}_{31}\times{\bf f}_3=0$.
Now we can eliminate the dynamics of the background fluid and study the dynamics of the swimmer. 
We make further mathematical simplification by assuming that the changes in arm lengths are small: 
$L_{i}\gg u_{1},u_{2},a\, (i=f,b)$. Throughout what follows, we will keep only the leading 
order terms in terms of these small quantities. One note that this Taylor expansion is performed to simplify the mathematical analytical results and 
it   does not change the generic features of the results.   
We can express the overall dynamics in terms of the internal motions given by $u_{1}(t)$ and $u_{2}(t)$.   The time averaged linear velocity of the swimmer's center up to the leading order of the 
small quantities $u_1/L$, $u_2/L$ and $a/L$ 
reads:
\be 
\langle\mathbf{V}_s\rangle\, _{t} = \dot{\gamma} \, z\, \hat{\mathbf{i}} + \left( V^{0} + V^{\dot{\gamma}} \right) \hat{\mathbf{t}}
\ee
Here, $V^{0}= - \frac{7}{24} \frac{a}{L^2} (1-\delta) \Phi$ is the swimming speed of a swimmer in a quiescent flow, and 
$V^{\dot{\gamma}} =-\frac{7}{4} a\,\dot{\gamma} \, \lambda \,\delta$ 
represents the 
velocity changes due to the presence of a shear flow.
Here $\Phi = \langle u_{1} {\dot{u}}_{2} - u_{2} {\dot{u}}_{1}\rangle\,_t$ and $\lambda = \sin\theta \cos\theta \cos\varphi$. We have further assumed that the asymmetry parameter $\delta$ is also very small. Externally applied shear flow, 
tends to rotate the swimmer. Angular velocity of the swimmer 
in polar coordinate can be written as: $\vec{\omega} = \dot{\theta} \, \hat{\boldsymbol{\theta}} + \dot{\varphi} \, 
\sin{\theta} \, \hat{\boldsymbol{\varphi}}$, with:
\be
\dot{\theta} = \dot{\gamma} \, \cos^2{\theta} \cos{\varphi},~~~
         \dot{\varphi} = - \dot{\gamma} \cot{\theta} \sin{\varphi}.
\ee

In addition to the swimming velocity, we can calculate the hydrodynamic forces exerted on fluid by the spheres. 
To express the forces, we decompose all forces as: 
${\bf f}_i={\bf f}_{i}^{0}+{\bf f}_{i}^{\dot{\gamma}}$ where the contribution due to the 
shear flow is explicitly separated from the terms that are present for swimmer moving in a 
quiescent flow.  Time averaged forces for a swimmer moving in a quiescent flow are given by:  
\begin{align}\nn
\langle {\bf f}^{0}_{2}\rangle\,_{t} &= -\frac{5}{8}\pi\eta_{0} {\left(\frac{a}{L}\right) }^2 
(1+\frac{7}{5}\delta) \Phi \,{\hat {\bf t}},      \nn\\
\langle {\bf f}^{0}_{3}\rangle\,_{t}&=-\frac{5}{8} \pi\eta_{0} {\left(\frac{a}{L}\right) }^2 (1-\frac{17}{5} \delta)\Phi \, {\hat {\bf t}},    \nn
\end{align}
and the contribution due to the shear flow are given by:
\begin{align}\nn
\langle {\bf f}^{{\dot \gamma}}_{2}\rangle\,_{t}& = - 6 \pi \eta_{0} \, a\, L \,  \lambda \, \dot{\gamma} \,(1+
\frac{1}{3}\delta)\, {\hat {\bf t}},  \nn\\
\langle {\bf f}^{{\dot \gamma}}_{3}\rangle\,_{t}& = 6 \pi \eta_{0} \, a\, L \,  \lambda \, \dot{\gamma} \, (1+
\frac{2}{3}\delta)\, {\hat {\bf t}}. \nn
\end{align}
We will use the above result to extract the rheological response of a fluid containing a collection of 
such swimmers. 
But before studying the properties of a suspension, we note that the force-dipole tensor for a swimmer moving in a 
quiescent flow can be written as: ${\bf D}=6\pi\eta_{0}(a/L)^2\delta L\Phi{\hat {\bf t}}{\hat {\bf t}}$. It is shown that far field velocity of a swimmer can be expanded on terms of the moments of the forces~\cite{Lauga}. As one can see for an asymmetric swimmer with $\delta\neq 0$, the force dipole velocity field dominates the far field behavior. An intuitive classification of the swimmers is based on the observation that how the driving force of the motion is located at the head or at the tail of the swimmer. 
For pushers the driving force sits on the tail 
while for pullers the driving force sits on the head. Referring to  Fig.~(\ref{fig1}), for $\Phi<0$, the intinsic velocity of the swimmer is in the direction given by $\hat{\mathbf{t}}$, and it is the sign of $\delta$ which determines whether the swimmer is puller or pusher. For $\delta>0 (<0)$ the swimmer is pusher (puller).

After calculating the forces exerted by the spheres to the fluid,  
 the contribution to the stress tensor of the fluid from a single swimmer can be obtained.  
For a dilute suspension of $N$ objects (swimmers) moving  in an ambient fluid with viscosity $\eta_{0}$ and volume $V$, the stress tensor averaged over the surface of objects  can be written as: $\sigma = \sigma^0+\frac{N}{V} \mathcal{S}$~\cite{Batchelor}, where $\sigma^0$ 
represents the stress tensor if all the objects are removed from the fluid.
The tensor $\mathcal{S}$ is  the extra stress due to the presence of a  single object and it reads:
\be \nn
\mathcal{S}=-\int \Big[ \frac{1}{2}\big( \mathbf{r} \boldsymbol{\sigma}\cdot \hat{\bf{n}}+\boldsymbol{\sigma}\cdot \hat{\bf{n}} \mathbf{r}\big) - \frac{1}{3}\big(\mathbf{r} \cdot\boldsymbol{\sigma}\cdot \hat{\bf{n}} \big) \bf{I}\Big] \,\mathrm{d}A
\ee
where the integral is over the surface of the body and ${\sigma}\cdot \hat{\bf{n}}$ is the stress at any point $\bf{r}$ on the body. In the present  case, each object is a collection of three connected 
spheres and in the limit of approximations performed before, they are replaced by 
point forces. In terms of the forces exerted by spheres, this stress tensor  reads: 
${\cal S}=-{\bf r}_{21}{\bf f}_2-{\bf r}_{31}{\bf f}_3$. Inserting the results given before and averaging over time we can keep the leading order terms:
\be\nn 
\langle\mathcal{S}\rangle_t =\displaystyle{ - \frac{29}{4} \pi \eta_{0} \,\frac{a^2}{L}  \, \Phi \,   \delta  \,\, \hat{\mathbf{t}} \, \hat{\mathbf{t}}  \, + \, 12 \pi \eta_{0} \, a L^2 \,\dot{\gamma}\, ( 1+\delta )  \,\lambda \, \hat{\mathbf{t}} \, \hat{\mathbf{t}} }                               
\label{stress2}
 \ee
The first part in the above stress, is independent of the shear rate (${\dot \gamma}$). 
We call this part as active contribution to the 
stress tensor, in the sense that it depends on 
the internal activity of the swimmer. The second part of the stress tensor reflects a shear rate dependent 
contribution that is a passive part. 
There is no cross term in the stress tensor with simultaneous  dependence  both on swimmer's activity and the shear rate. This is due to the linearity of the Stokes equation that does not allow any direct combination between the external stress and the internal activity of the swimmer. 
As one can distinguish, the stress contribution from a swimmer depends crucially on the angular direction of the swimmer. Assuming that our system contains a collection of randomly orientated swimmers, we can average the stress tensor over all swimmers to reach the following 
result for the off-diagonal elements of the total stress tensor:
$$\sigma_{ij}=\frac{\eta^{\text{eff}}}{2}(\delta_{i,x}\delta_{j,z}+\delta_{j,x}\delta_{i,z})$$
Comparing this result with the corresponding components of stress tensor for a simple shear flow, we 
can easily conclude that a dilute suspension of swimmers behaves like a Newtonian fluid where 
its viscosity is replaced by $\eta^{\text{eff}}$. Effective viscosity of a homogeneous and dilute suspension of  swimmers is given by:
$$\eta^{\text{eff}}=\eta_{0}\left(1+  \frac{5}{2} c\, (\frac{L}{5a})^{2}\right),$$
where 
$c=4\pi a^3 \frac{N}{V}$ is the volume fraction of $N$ swimmers distributed in a fluid with volume $V$. 
This result is comparable 
with the Einstein's relation for the effective viscosity of a dilute suspension of objects  each has  volume $V$. Please note that the contribution from swimmer's activity (the active part in the stress tensor) averages to zero and it does not have any influence in the effective viscosity of the suspension. This is understandable in the sense that the viscosity measures the linear response of the fluid to an externally applied shear and the active term in the stress does not depend on the shear rate at all. Appearing any cross term in the stress tensor would cause the suspension to have effective viscosity proportional to the activity of swimmers. But here, linearity of the Stokes equation does not allow any such contribution. Please 
note that this point of view is in contrast with the phenomenology that is used in previous 
works \cite{Haines}. 

Our central question in this work is the influence of swimmers activity to the fluid rheological properties. Here 
we will show, taking into account the thermal fluctuations of the swimmers introduce a mechanism that can  couple the internal activity of the swimmer to the externally applied shear rate. This will result a non trivial 
non-Newtonian behavior for the fluid. To study the thermal fluctuations of the swimmers, we denote the angular distributions of such swimmers by $F(\theta,\varphi)$. In addition to time averaging over the internal motion of an individual swimmer, we need to average over the angular distribution as well. Having in  hand the distribution function, we can perform the averaging procedure for a function $\delta {\cal S}$ as below:
\be
\langle\delta {\cal S}\rangle_{t,e}=\langle\int d\Omega \, F(\theta,\varphi)\, \delta {\cal S}\rangle_{t},
\label{averaging}
\ee
where the integral should be taken over the polar angles. Please note that 
we first average  over an ensemble of swimmers with instantaneous same internal configuration then average over time. Validity of the above order of averaging will be discussed later. The distribution function obeys the following Fokker-Planck equation:    
\be 
\frac{\partial F}{\partial t}= D_{r} {\nabla}^2 F - \nabla \cdot (F \,\vec{\omega}),
\ee
where ${\nabla}^2$ is the angular part of the Laplacian operator and the velocity in the angular space is given by:  
 $\vec{\omega} = \dot{\theta} \,\hat{\boldsymbol{\theta}} + \dot{\varphi} \, \sin{\theta} \,\hat{\boldsymbol{\varphi}}$. Here $D_{r}$ (in ${\mathrm{sec}}^{-1}$) stands for the rotational diffusion coefficient of the swimmers and it depends both on thermal energy $k_BT$ and the rotational friction coefficient of the swimmer $\xi_r$ by $D_r=k_BT/\xi_r$. 
Rotational friction coefficient is a hydrodynamical quantity that measures the response of the swimmer to an externally applied torque. Direct calculations for a three-sphere system immersed in a quiescent fluid and subject to external torque, reveals that: $\xi_r=4\pi\eta_{0} a(L_b^2+L_f^2+L_bL_f)$. In terms of small dimensionless Peclet number $ \mathsf{Pe} :=(\dot{\gamma}/k_BT)\xi_r$, we can set up a perturbation steady state solution for the above Fokker-Planck equation as: 
\begin{align} \nn
 &  F(\theta , \varphi)= \frac{1}{4\pi} \Big[ 1+ \frac{ \mathsf{Pe}}{2} \big( \sin{2\theta} \cos{\varphi}  \big) + 
 \frac{{ \mathsf{Pe}}^2}{280} \bigg( \big(  7 - 35 {\cos}^4{\theta} \big)          \nn \\     
 & \qquad  +\big(  35 \,{\sin}^2{\theta} \,{\cos}^2{\theta} + 5 \, \, {\sin}^2{\theta}\big) \cos{2\varphi} \bigg)+ \cdots  \Big].  
\end{align}
The Peclet number is essentially the ratio between the diffusion time $\tau_D=D_r^{-1}$ and the time scale of the applied shear flow $\tau_s={\dot \gamma}^{-1}$. Using this distribution function we can calculate all statistical variables up to the second order of small parameter 
$\mathsf{Pe}$. To obtain the rheological properties of the suspension we should  repeat the  calculations similar to what have done before Eq.\ref{stress2}. Along such calculations we should perform an ensemble average followed by time average over internal configuration. We will need the following ensemble averages: 
\be \nn
 \displaystyle{  \langle\hat{\mathbf{t}}\, \hat{\mathbf{t}}\rangle_{t,e} = \left( \begin{array}{ccc}
 \frac{1}{3} + \frac{1}{105} { \mathsf{Pe}}^2 &0 &\frac{1}{15}  \mathsf{Pe}  \\
 0 & - \frac{1}{120} { \mathsf{Pe}}^2 &0  \\
 \frac{1}{15}  \mathsf{Pe}  & 0 &\frac{1}{3} - \frac{23}{840} { \mathsf{Pe}}^2 \\
\end{array}\right) + \mathcal{O}({ \mathsf{Pe}}^3)},
\ee
and
\be \nn
 \displaystyle{  \langle \lambda\hat{\mathbf{t}}\, \hat{\mathbf{t}}\rangle_{t,e} = \left( \begin{array}{ccc}
 \frac{1}{35}  \mathsf{Pe} &0 &\frac{1}{15} +\frac{52}{315} { \mathsf{Pe}}^2 \\
 0 & \frac{1}{105}  \mathsf{Pe} &0  \\
 \frac{1}{15} +\frac{52}{315} { \mathsf{Pe}}^2  & 0 & \frac{1}{35}  \mathsf{Pe} \\
\end{array}\right) + \mathcal{O}({ \mathsf{Pe}}^3)}.
\ee
Using the above results, it is  straightforward to  calculate  the rheological properties of the dilute suspension of swimmers. Effective viscosity of the suspension is the first and most important quantity 
that we can address here. For a dilute suspension of swimmers with volume fraction $c$, that was defined before, the effective viscosity reads: 
\be 
{\eta}^{\text{eff}} = {\eta}_{0}\Big[ 1+ c \Big( \frac{5}{2} {(\frac{L}{5 a})}^{2} - \frac{W}{k_{B} T}\delta(\frac{L}{a})   \Big)  \Big],
\label{effective2}
\ee
where $W=\frac{213\pi}{14} {\eta}_{0} \, L^2 \,\, \bar{v}$, and $\bar{v}=- \frac{7}{24} \frac{a}{L^2}\Phi$.
At very high temperature condition, where the effects of thermal fluctuations dominates, we can recover the 
previous Einstein's like behavior for the viscosity of suspension. 
The main result that we  obtain here is the influence of swimmer's activity in the viscosity. 
The activity of the swimmer contributes in the viscosity through  function $W$. 
Asymmetry parameter $\delta$ appears in the above result ; this is another important feature of our results which reflects the fact that viscosity for pushers or pullers are different. 
For pushers (pullers), $\delta>0$ ($\delta<0$) and one can easily see 
that $\eta^{\text{eff}}<\eta_{0}$ ($\eta^{\text{eff}}>\eta_{0}$). 
Such viscosity reduction and enhancement for a suspension of active particles are in clear 
agreement with the  experiments have been performed recently. 
In addition to the effective viscosity of the suspension, the fluctuation of the active particles 
will mediate an asymmetric behavior for the fluid. Normal-stress differences quantify this 
non-Newtonian behavior of the suspension. Normal-stress differences, up to second order in Peclet number, reads: 
\begin{align}\nn
&N_{1}= {\sigma}_{xx} - {\sigma}_{yy} =  - W' ( \frac{1}{3}+ \frac{1}{56}{\mathsf{Pe}}^2 )  - \frac{8}{35} (\pi \eta_{0} a L^2 \dot{\gamma} ) \mathsf{Pe} \nn \\
&N_{2}= {\sigma}_{xx} - {\sigma}_{zz} = - \frac{31}{840} W'  {\mathsf{Pe}}^2 ,\nn
\end{align}
where $W' = (a/L) \delta \,W$. Note that  these normal stress differences have different signs for pushers and pullers. This may reflect itself in standard rheological experiments, for example in Weisenberg effect. 

To have an intuition about the mathematical results we have obtained so far, we can think about the numerical values of the parameters. Diffusion time scale $\tau_D$,  time scale for shear flow $\tau_s$ and the 
internal time scale for an individual swimmer $\tau_i=u_0^2/|\Phi|$, are three different time scales that characterize our system. $\tau_D$ is a passive parameters that crucially depends on the size of the swimmer. 
For a swimmer with the largest length $L\sim 1\mu \mathrm{m}$ and the smallest 
length $a\sim 0.1\mu \mathrm{m}$, we see that $\tau_D\sim 1 \mathrm{sec}$. The 
validity of the averaging order considered in Eq.\ref{averaging}, can be satisfied by $\tau_i\gg\tau_D$. 
For a real microswimmer, for example Spermatozoa, the undulation frequency of flagellum that is about $10\mathrm{Hz}$, sets the internal time scale that is greater than the diffusion time. 
How important is the activity contribution to the effective viscosity? To answer this question, one can see from Eq.\ref{effective2}, that the ratio between the active and passive correction to the viscosity is given by: $\delta(a/L)(W/k_BT)$. Assuming that $\delta\sim a/L \sim 0.1$ and at room temperature, for a typical swimmer with swimming velocity $\bar{v}\sim 1\mu \mathrm{m}/\mathrm{sec}$, the above ratio 
is of the same order as the passive part. 

To summarize, we have shown how a collection of  thermally  fluctuating swimmers can mediate 
an effective viscosity for the suspension. Depending on the type of individual swimmers, whether 
those are pushers or pullers, the effective viscosity would be smaller or larger than the viscosity 
of the ambient fluid. In addition to the effective viscosity, the normal stress differences are also calculated for such suspension.

A natural extension of our work is the generalization to the systems with larger volume fractions. 
Active and passive hydrodynamic interactions between swimmers \cite{Yeomans,Farzin,Najafi3} may influence the 
viscosity for systems that are beyond the dilute regime considered here.   
One may imagine that in a suspension of  microorganisms, another mechanism related to the  internal feedback of the organism may provide additional source of viscosity change. If we assume that an  organism can changes its internal undulation proportional to the rate of external shear, this scenario also will mediate an effective contribution to the viscosity.

\bibliography{3SNN}




\end{document}